\pdfoutput=1
\documentclass[aps,twocolumn,floatfix,prl]{revtex4}
\usepackage{graphicx,amsmath,bbm}

\newcommand{\ua}{\uparrow}
\newcommand{\da}{\downarrow}

\begin{document}

\title{Non-Gaussian fluctuations of mesoscopic persistent currents}
\date{\today}

\author{J.\ Danon}
\author{P.~W.\ Brouwer}
\affiliation{Dahlem Center for Complex Quantum Systems, Freie Universit\"{a}t Berlin, Arnimallee 14, 14195 Berlin, Germany}

\begin{abstract}
The persistent current in an ensemble of normal-metal rings shows Gaussian distributed sample-to-sample fluctuations with non-Gaussian corrections, which are precursors of the transition into the Anderson localized regime. We here report a calculation of the leading non-Gaussian correction to the current autocorrelation function, which is of third order in the current. Although the third-order correlation function is small, inversely proportional to the dimensionless conductance $g$ of the ring, the mere fact that it is nonzero is remarkable, since it is an odd moment of the current distribution.
\end{abstract}

\maketitle

Already in the early days of quantum mechanics, it was realized that
the ground state of small conducting ring-like structures penetrated
by a magnetic field could contain a {\it dissipationless} current
circling through the structure~\cite{hund:annphys}. Such persistent
currents have been well known in the context of superconducting
devices since the early 1960s~\cite{Tinkham}. Normal-state persistent
currents became the subject of intensive research only after it was
understood that elastic impurity scattering, unavoidable in almost all 
realizations, reduces the phenomenon, but does not completely suppress it~\cite{buttiker:physlett}:
The typical magnitude of a persistent
current in a disordered normal-metal ring with circumference $L$ and
diffusion constant $D$ is $I \sim e/\tau_{L}$, where $\tau_{L} =
L^2/D$ is the time it takes an electron to diffuse once around the ring
\cite{PhysRevLett.62.587}.

Because of time-reversal symmetry, a persistent current must vanish at
zero field, whereas gauge invariance imposes that the current is
periodic in the flux $\phi$ penetrating the ring, with period $\phi_0
= h/e$~\cite{PhysRevLett.7.46}. The precise value of the current $I(\phi)$
depends strongly on the specific disorder realization. When averaging
over disorder, the odd Fourier components of the average current
$\langle I(\phi) \rangle$ become vanishingly small.
As long as the applied magnetic field is weak (such that
time-reversal symmetry is preserved inside the metal ring), the even 
Fourier components 
of $\langle I(\phi)\rangle$ are nonzero, but still small compared to those of the typical current $\langle I^2(\phi) \rangle^{1/2}$~\cite{PhysRevB.47.15449,PhysRevLett.65.381,PhysRevLett.66.88}.
Whereas early experiments reported a persistent current a few orders of magnitude larger than the theoretical prediction \cite{PhysRevLett.64.2074,PhysRevLett.67.3578}, improved SQUID detection of the magnetic response of
a small number of gold rings~\cite{PhysRevLett.102.136802} and, very 
recently, current detection through the frequency shift of cantilevers covered with 
aluminum rings~\cite{science:bleszynski} unambiguously
confirmed the predictions for the average and typical current
amplitudes.

The sample-to-sample distribution of $I(\phi)$ is well approximated by
a Gaussian if the dimensionless conductance $g$ of the rings is large,
but deviations from Gaussian statistics are expected to become
relevant if the rings are very thin or strongly disordered, in which
case $g$ is small. The unprecedented accuracy of the cantilever method
opens the possibility to measure the full probability distribution of
the current, not just its first two moments. 
Non-Gaussian fluctuations play an important role in theoretical
and experimental studies of transport in disordered metals
\cite{PhysRevB.55.4710,PhysRevLett.88.146601}, being a
precursor of the transition to the Anderson localized regime. Non-Gaussian fluctuations were also considered in the conductance distribution of chaotic quantum dots, for which they are an indicator of fully quantum coherent 
transport \cite{huibers1998}.

In this Letter, we calculate the leading non-Gaussian connected autocorrelation
function 
\begin{equation}
K(\phi_1,\phi_2,\phi_3) = \langle I(\phi_1) I(\phi_2)
I(\phi_3) \rangle_{\rm c}
\end{equation}
of persistent currents to leading order in
$1/g$. Here the brackets $\langle \ldots \rangle$ denote a disorder
average and the subscript `c' refers to the `connected average',
$\langle abc \rangle_{\rm c} = \langle abc \rangle - \langle ab
\rangle \langle c \rangle - \langle ac \rangle \langle b \rangle -
\langle bc \rangle \langle a \rangle + 2 \langle a \rangle \langle b
\rangle \langle c \rangle$. At equal values of the arguments,
$K$ gives the third cumulant $\langle I(\phi)^3 \rangle_{\rm c}$
of the current distribution, which is proportional to its skewness. 
As we will show below, $\langle I(\phi)^3 \rangle_{\rm c}$ is nonzero and
of order $\sim e^3/g \tau_{L}^3$ for small magnetic fields only, and
becomes vanishingly small for large magnetic fields. However, the
full third-order correlation function $K(\phi_1,\phi_2,\phi_3)$ remains 
nonzero for arbitrary magnetic field strengths. This is remarkable 
because $K(\phi_1,\phi_2,\phi_3)$ addresses an odd moment of the 
current distribution.
Previously published order-of-magnitude estimates for $K$~\cite{smithambegaokar,eckernschmid} are a factor $\sim \tau_L/\tau$ larger than the result of our calculation, $\tau$ being the elastic mean free time~\cite{footnote}.

We will first describe the calculation of the connected
autocorrelation function $K(\phi_1,\phi_2,\phi_3)$ for zero
temperature, at weak magnetic fields, and in the absence of spin-orbit
scattering. Then we evaluate the effect of finite temperature,
larger magnetic fields, and significant spin-orbit scattering. Finally, we also
discuss the possibility to identify the non-Gaussian current
fluctuations in persistent current measurements.

Starting point of a calculation of the persistent current is the thermodynamic 
relation~\cite{PhysRevLett.66.88,PhysRevLett.70.1976}
\begin{equation}
I = -\frac{\partial}{\partial \phi}\left\{\Omega(\mu,\phi) + \frac{1}{2
    \langle \nu \rangle}
  (N(\mu,\phi) - \langle N \rangle)^2 \right\},
\label{eq:eq1}
\end{equation}
which expresses the current $I$ in a ring with a fixed number of
electrons in terms of the grand canonical potential $\Omega$ and the
grand canonical fluctuations of the particle-number $N$. Both $\Omega$
and $N$ are calculated as integrals of the density of states $\nu(E,\phi)$,
\begin{equation}\begin{split}
\Omega(\mu,\phi) & = - k_B T \int dE \,\,\nu (E,\phi) \ln [ 1+e^{-(E-\mu)/k_B T}],\\
N(\mu,\phi) & = \int dE \,\,\frac{\nu(E,\phi)}{1+e^{(E-\mu)/k_B T}},
\end{split}
\label{eq:eq2}
\end{equation}
where $T$ is the temperature and $\mu$ the
chemical potential, which is chosen such that the disorder average
$\langle N(\mu)\rangle$ equals the (canonical) particle number
$N$.

\begin{figure}[t]
\begin{center}
\includegraphics[width=8.5cm]{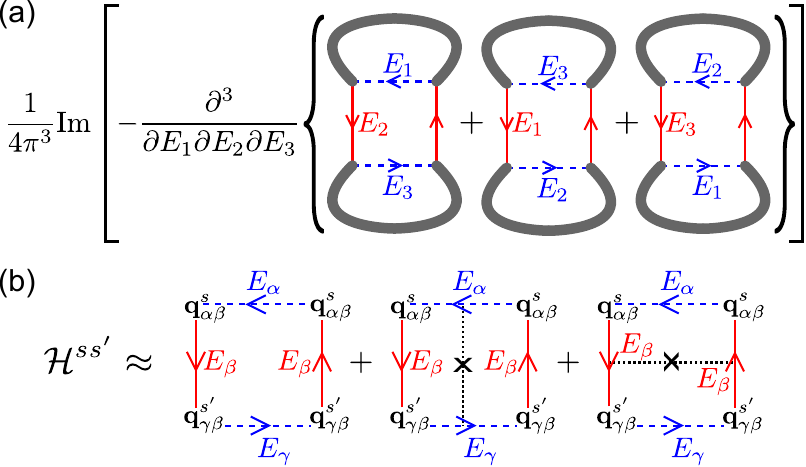}
\caption{(Color online.) Diagrammatic representation of (a) the third-order autocorrelation function of the density of states, $\langle \nu(E_1,\phi_1) \nu(E_2,\phi_2) \nu(E_3,\phi_3) \rangle_c$ and of (b) the Hikami box $\mathcal{H}$. The advanced (retarded) Green functions are represented by blue dashed (red solid) lines. The thick gray lines connecting the Green functions represent the diffuson and cooperon ladders. The dotted black lines represent impurity scattering.}\label{fig:fig1}
\end{center}
\end{figure}

Since $\Omega(\mu,\phi)$ and $N(\mu,\phi)$ are statistically
uncorrelated to leading order in $1/g$ --- they have opposite parity 
under a particle-hole conjugation ---, only the first term in Eq.\
(\ref{eq:eq1}) contributes to the autocorrelation function
$K$. Calculation of $K$ then proceeds through the standard relation 
between the density of states $\nu(E,\phi)$ and the Green function,
\begin{equation}
  \nu(E) = \frac{1}{2 \pi i} \sum_{{\bf q},\sigma}
  [G^{\rm A}_{\sigma}({\bf q},{\bf q};E) - G^{\rm R}_{\sigma}({\bf q},{\bf q};E)],
  \label{eq:nu}
\end{equation}
where $\sigma =\ua,\da$ is the spin index.

Calculation of the current autocorrelation function $K$ requires calculation of the connected density-of-states autocorrelation function $\langle \nu(E_1,\phi_1) \nu(E_2,\phi_2) \nu(E_3,\phi_3) \rangle_c$.
Performing the disorder average using diagrammatic perturbation theory~\cite{akkermans},
the latter
can be represented
schematically as in Fig.\ \ref{fig:fig1}a. In this figure, the 
advanced (blue dashed lines) and retarded (red solid lines) Green 
functions are connected to each other by thick gray lines which represent 
the diffuson ($-$) and cooperon ($+$) ladders, 
$\mathcal{D}^\pm_{\sigma_{\alpha},\sigma_{\beta}}
({\bf q}_{\alpha\beta}^{\pm};E_{\alpha\beta})$,
where we use the short-hand notations 
${\bf q}^\pm_{\alpha\beta} = {\bf q}_{\alpha} \pm {\bf q}_{\beta}$ and 
$E_{\alpha\beta} = E_{\alpha} - E_{\beta}$. In this notation, the index $\alpha$ refers to the wave vector, energy and spin of the advanced Green function and $\beta$ to those of the retarded function. In the absence of a magnetic field penetrating the metal ring and without spin-orbit scattering, the diffuson and cooperon 
propagators are given by the standard expression
\begin{eqnarray}\label{eq:D}
\mathcal{D}^\pm_{\sigma_{\alpha},\sigma_{\beta}}({\bf q}_{\alpha\beta}^{\pm};E_{\alpha\beta}) &=&
\mathcal{D}^\pm({\bf q}_{\alpha\beta}^{\pm};E_{\alpha\beta})  
  \nonumber \\ &=&
\frac{\hbar}{\pi\langle\nu\rangle \tau^2 }\frac{1}{D |{\bf q}_{\alpha\beta}^{\pm}|^2 + \frac{i}{\hbar} E_{\alpha\beta}}.
\end{eqnarray}
The ladders are connected at a `Hikami box' $\mathcal{H}$ depicted in Fig.\ \ref{fig:fig1}b, 
\begin{equation}
\mathcal{H}^{ss'}
  = \frac{\pi \langle \nu \rangle\tau^4  }{\hbar^3}
  \left[D|{\bf q}_{\alpha\beta}^{s} -{\bf
  q}_{\gamma\beta}^{s'}|^2+\frac{i}{\hbar}(
  E_{\alpha\beta} + E_{\gamma\beta}) \right],
\label{eq:eq5}
\end{equation}
where the indices $s$ and $s'$ can be $+$ or $-$, depending on whether the upper ($s$) or lower ($s'$) ladders represent diffuson or cooperon propagators.
Combining the expressions for $\mathcal{D}$ and $\mathcal{H}$, one then computes the third-order density-of-states autocorrelation function by summing over the momenta ${\bf q}_{1}$, ${\bf q}_{2}$, and ${\bf q}_{3}$, the corresponding spin indices, and the diffuson/cooperon indices $s$ and $s'$.

We now take the ring's circumference $L$ to be much larger than its
thickness. In that limit, only terms in which the transverse component 
of the momenta ${\bf q}^{\pm}$ vanishes contribute to the correlation
function. In the presence of a flux $\phi_\alpha$ penetrating the ring, the longitudinal component $q$ of 
the momentum takes the discrete values
\begin{equation}
  q_{\alpha} = \frac{2 \pi}{L} (n - \varphi_{\alpha}), \quad\text{with}\quad
  \varphi_{\alpha} = \frac{\phi_{\alpha}}{\phi_0},
\end{equation}
$n$ being an integer. Converting the sums over the integers $n$ to integrals using the Poisson summation rule, we then find that, at temperature $T=0$,
\begin{widetext}
\begin{equation}
  K(\phi_1,\phi_2,\phi_3) = \frac{12 e^3}{\pi^6  \tau_L^3 g}
  \frac{\partial}{\partial \varphi_1}
  \frac{\partial}{\partial \varphi_2}
  \frac{\partial}{\partial \varphi_2}
  \sum_{\alpha,\beta,\gamma=1}^3 \sum_{s,s' = \pm} \sum_{k,k'=1}^{\infty}
  \frac{k^2+4kk'+k'^2}{(kk')^2(k+k')^4} 
  \sin(2 \pi k \varphi_{\alpha\beta}^{s}) 
  \sin(2 \pi k' \varphi_{\gamma\beta}^{s'}),
\label{eq:eq7}
\end{equation}
\end{widetext}
where the dimensionless conductance $g = 2\pi\hbar\langle\nu\rangle / \tau_L$, and with the notation $\varphi^\pm_{\alpha\beta} = \varphi_\alpha \pm \varphi_\beta$.
The scale of the non-Gaussian fluctuations, when
normalized to the Gaussian fluctuations, is 
$K^{1/3} / \langle I^2\rangle^{1/2}\sim g^{-1/3}$.
The magnitude of the summand in Eq.\ (\ref{eq:eq7})
drops to zero quickly for increasing 
$k$ and $k'$, so that only the lowest harmonics will contribute 
significantly to $K$. Note that the third 
cumulant,
\begin{equation}
\begin{split}
  \langle I(\phi)^3 \rangle_{\rm c} &= -
  \frac{1152 e^3}{\pi^3 \tau_{L}^3 g}
  \sum_{k,k'=1}^{\infty}
  \frac{k^2+4kk'+k'^2}{k'(k+k')^4} \\
  & \qquad\qquad\qquad \times \sin(4 \pi k \varphi) \cos(4 \pi k' \varphi),
\end{split}
  \label{eq:7b}
\end{equation}
is non-zero only due to the 
`double cooperon' contribution with $s=s'=+$.

Let us now discuss the effect of a finite magnetic field, spin, and finite
temperature. All three effects are relevant for a realistic description of 
the experiment of Ref.~\cite{science:bleszynski}. Following 
Ref.~\cite{PhysRevB.81.155448}, we model the magnetic field penetrating 
the sample as toroidal. Although a toroidal magnetic field 
does not fully 
correspond to the experimental geometry, it is believed to provide a 
reliable and tractable description of the experiment 
\cite{PhysRevB.81.155448}. With a toroidal magnetic field $B_{\parallel}$, 
the transverse component of the momenta ${\bf q}^{\pm}$ is no longer 
zero, which leads to a shift of the momentum eigenvalues appearing 
in the expression for the cooperon ladders,
\begin{equation}
  (n-\varphi^+)^2 \to (n-\varphi^+)^2 + \epsilon_\bot^2,
  \label{eq:eq9}
\end{equation}
$\epsilon_{\bot}$ being a geometry-dependent positive constant 
proportional to $B_{\parallel}$.
For a ring with circular cross section with radius $R$, 
one has $\epsilon_\bot^2 = (1/8\pi^2h^2)(eLRB_\parallel)^2 + \mathcal{O}[B_\parallel^4]$ 
\cite{PhysRevB.81.155448}.

When the coupling of the applied magnetic field to the electron spin (with 
Zeeman frequency $\omega_{\rm Z}$) or spin-orbit scattering (with spin-orbit 
scattering time $\tau_{\rm so}$) are included, the diffuson and cooperon 
propagators acquire additional structure with respect to their spin indices. 
As long as 
$\tau/\tau_{\rm so} \ll 1$,
the Hikami box remains spin-conserving, 
such that elements of the propagators
which flip the spin of one of the Green functions do not contribute.
In this case only `diagonal' elements play a role and the two spin indices
in Eq.\ (\ref{eq:D}) continue to be sufficient.
With the toroidal field (\ref{eq:eq9}) included,
the spin-dependent propagators read
\begin{equation}
\begin{split}
\mathcal{D}^{\pm}_{\ua,\ua}({\bf q}^{\pm},E) &  = \mathcal{D}_{\da,\da}^{\pm}({\bf q}^{\pm},E) \\ & = \frac{1}{2} \sum_{c=\pm 1} \mathcal{D}^{\pm}({\bf q}^{\pm},E-\frac{i\hbar}{\tau_L}x^\pm_{1,0,c})\\
\mathcal{D}_{\ua,\da}^{\pm}({\bf q}^{\pm},E) &  = \frac{1}{2}\sum_{c=\pm 1}\mathcal{D}^{\pm}({\bf q}^{\pm},E - \frac{i\hbar}{\tau_L}x^\pm_{-1,1,c})\\
\mathcal{D}_{\da,\ua}^{\pm}({\bf q}^{\pm},E) &  = \frac{1}{2}\sum_{c=\pm 1}\mathcal{D}^{\pm}({\bf q}^{\pm},E - \frac{i\hbar}{\tau_L}x^\pm_{1,1,c}),\\
\end{split}
\label{eq:eq10}
\end{equation}
where ${\mathcal D}^{\pm}({\bf q}^{\pm},E)$ is the spin-independent propagator of Eq.\ (\ref{eq:D}), and the factors $x^\pm_{a,b,c}$ are defined as
\begin{equation}
\begin{split}
x^-_{a,b,c} & = \frac{2\tau_L}{3\tau_\text{so}}[1+c+b(1-c)]+ab(i\tau_L\omega_z) \\
x^+_{a,b,c} & = \frac{2\tau_L}{3\tau_\text{so}}(2-b)\\
	& \qquad+bc\frac{\tau_L}{3\tau_\text{so}}\sqrt{4-9(\tau_\text{so}\omega_Z)^2} + 4\pi^2\epsilon_\bot^2.
\end{split}
\label{eq:xym}
\end{equation}
For a large Zeeman splitting ($\tau_L\omega_Z \gg 1$), the 
contribution of the elements $\mathcal{D}_{\ua,\da}$ and 
$\mathcal{D}_{\da,\ua}$ will be suppressed, which leads 
to a factor four decrease of $K(\phi_1,\phi_2,\phi_3)$ with respect to the 
zero-field case of Eq.\ (\ref{eq:eq7}). Strong spin-orbit scattering 
($\tau_L/\tau_{\rm so} \gg 1$) causes an additional factor four decrease of 
$K(\phi_1,\phi_2,\phi_3)$.

\begin{figure}[b!]
\begin{center}
\includegraphics[width=8.5cm]{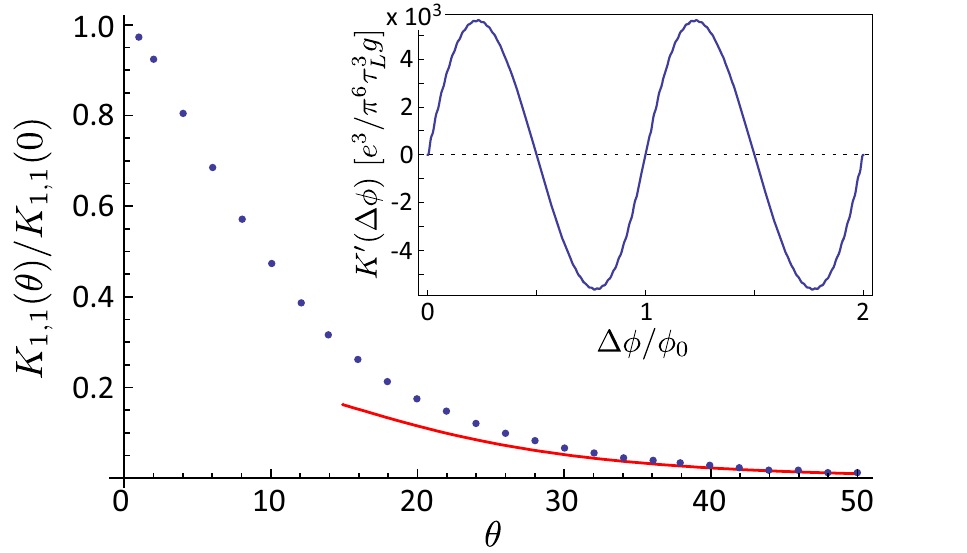}
\caption{(Color online.) (Main) The Fourier component $K_{1,1}$ as a function of the rescaled temperature $\theta = 2\pi k_BT\tau_L/\hbar$. $K_{1,1}$ is normalized to the zero-temperature value predicted by Eq.\ \ref{eq:eq7}. For $\theta\gtrsim 25$ the high-temperature approximation given by Eq.\ \ref{eq:KTemp} (red solid line) and a numerical evaluation using Eq.\ \ref{eq:temp} (circles) agree well. (Inset) The correlator $K(\phi,\phi,\phi+\Delta\phi)$ at zero temperature in units of $e^3/\pi^3\tau_L^3g$. For this plot we used Eq.\ \ref{eq:eq7}, summing over $k,k'\leq 40$.}
\label{fig:fig2}
\end{center}
\end{figure}

Inserting the diffuson and cooperon propagators of Eq.\ (\ref{eq:eq10}) and extending the previous calculation to finite temperatures, we arrive at the complete result
%
\begin{widetext}
\begin{equation}\label{eq:temp}
K(\phi_1,\phi_2,\phi_3) = \frac{e^3\theta^3}{64\pi^6\tau_L^3g}
  \frac{\partial}{\partial \varphi_{1}}\frac{\partial}{\partial
  \varphi_{2}}\frac{\partial}{\partial \varphi_{3}}
  \sum_{q=0}^{\infty} \sum_{\beta=1}^{3}
  \sum_{a=\pm} \text{Re}
  \left[ \Big\{ \sum_{p,k=1}^{\infty} \sum_{\alpha=1}^{3}
  \sum_{\substack{s,c = \pm}}\sum_{b=0}^{1}
  \sin(2 \pi k \varphi_{\alpha\beta}^{s})
  e^{- k\sqrt{x^{s}_{a,b,c}+\theta(p+q)}}\Big\}^2\right],
\end{equation}
where $\theta \equiv 2\pi k_BT\tau_L/\hbar$ is the rescaled temperature.

Equation (\ref{eq:temp}) shows explicitly how the various contributions to $K$ are suppressed by a magnetic field, spin-orbit scattering, and finite temperature. Using characteristic values for the experiment of Ref.\ \cite{science:bleszynski}, $L\sim 2.5~\mu$m, $R\sim 50$~nm and $B_\parallel\sim 3.5$~T, one finds that $\epsilon_\bot \sim 12$. Since all $x^+_{a,b,c}$ contain a term $4\pi^2\epsilon_\bot^2$, a large magnetic field, $\epsilon_\bot^2 \gg 1$, annihilates the contribution of all cooperons, and thereby also the third cumulant of Eq.\ (\ref{eq:7b}). In this regime, one thus indeed expects to find $\langle I(\phi)^3 \rangle_{\rm c} \approx 0$. However, since $x^-_{\pm 1,0,-1} = 0$ irrespective of the magnetic field and spin-orbit scattering strengths, the general connected correlation function remains nonzero at high fields and with strong spin-orbit scattering because of the contribution from the diffusons $\mathcal{D}^-_{\ua,\ua}$ and $\mathcal{D}^-_{\da,\da}$.

A large temperature leads to exponential suppression of all contributions to the connected correlation function $K(\phi_1,\phi_2,\phi_2)$ and to a suppression of the non-Gaussian fluctuations in comparison to the Gaussian fluctuations. In the high-temperature limit, $\theta \gg 1$, and for small magnetic fields and weak spin-orbit scattering, $\tau_L/\tau_{\rm so},\tau_{L}\omega_{Z} \ll 1$, (but still assuming $\epsilon_\bot^2 \gg 1$), 
Eq.\ (\ref{eq:temp}) simplifies to
\begin{equation}
  K(\phi_1,\phi_2,\phi_3) = \frac{8e^3\theta^3}{\pi^3\tau_L^3g}e^{-2\sqrt{\theta}}\Big\{ \sin[2\pi(\varphi_1+\varphi_2-2\varphi_3)]+\sin[2\pi(\varphi_2+\varphi_3-2\varphi_1)]+\sin[2\pi(\varphi_3+\varphi_1-2\varphi_2)] \Big\}.
  \label{eq:KTemp}
\end{equation}
\end{widetext}
For comparison, in the same limit the second-order current correlator, which describes the Gaussian fluctuations, scales as $\langle I^2 \rangle \propto \theta^2 e^{-\sqrt{\theta}}$~\cite{PhysRevB.47.15449}, so that typically $K^{1/3}/\langle I^2 \rangle^{1/2} \sim e^{-\sqrt{\theta}/6}g^{-1/3}$. In order to investigate the full temperature dependence of $K$, we show in Fig.\ \ref{fig:fig2} the temperature dependence of the leading Fourier component $K_{1,1}$ (which is the one given in Eq.\ (\ref{eq:KTemp}) above).

As a complete measurement of the connected correlation function $K(\phi_1,\phi_2,\phi_3)$ is likely to be very involved, it is worthwhile to investigate the behavior of the correlation function when two of its arguments are equal, $K(\phi,\phi,\phi+\Delta\phi)$. In the same limit of small magnetic fields and weak spin-orbit scattering, but with $\epsilon_\bot^2 \gg 1$, this correlator is a function only of the flux difference, $K(\phi,\phi,\phi+\Delta\phi) = K'(\Delta\phi)$. The zero temperature limit for $K'(\Delta\phi)$ is shown in the inset of Fig.\ \ref{fig:fig2}. In the experiment of Ref.\ \cite{science:bleszynski} one has roughly $g\sim 10^4$ for the smallest rings, so that $\sqrt[3]{K'(\Delta\phi)} \sim 0.1$~nA. We thus believe that a measurement of $K'(\Delta \phi)$ should be feasible. 

In conclusion, we have evaluated the third-order autocorrelation function for the persistent current in an ensemble of disordered metal rings. Remarkably, this correlator remains non-zero at large magnetic fields, although the non-Gaussian current fluctuations are a factor $\sim g^{1/3}$ smaller than the leading Gaussian fluctuations. The third-order connected correlator is the leading-order signature of localization effects. About one decade ago, the observation of non-Gaussian correlations in the transport of disordered quantum wires and chaotic dots was greeted as a unequivocal signature of quantum coherent transport \cite{PhysRevB.55.4710,PhysRevLett.88.146601,huibers1998}. We hope that our calculations motivate for a corresponding breakthrough in the realm of persistent currents and other equilibrium properties. 

The authors gratefully acknowledge helpful discussions with Teemu Ojanen and Felix von Oppen. This work is supported by the Alexander von Humboldt Foundation.


\begin{thebibliography}{10}
\bibitem{footnote}
Ref.~\cite{smithambegaokar} essentially presents a correct estimate, the disagreement stemming from a mistake made while combining intermediate results. In Ref.~\cite{eckernschmid} a Hikami cancellation was overlooked, giving rise to the extra factor $\tau_L/\tau$.
\end{thebibliography}

\end{document}